# Market-Based "Actual" Returns of Investors

Victor Olkhov

Independent, Moscow, Russia

victor.olkhov@gmail.com

ORCID: 0000-0003-0944-5113

Vers. 20 Feb., 2024

## ABSTRACT

We describe how the market-based average and volatility of the "actual" return, which the investors gain within their market sales, depend on the statistical moments, volatilities, and correlations of the current and past market trade values. We describe three successive approximations. First, we derive the dependence of the market-based average and volatility of a single sale return on market trade statistical moments determined by multiple purchases in the past. Then, we describe the dependence of average and volatility of return that a single investor gains during the "trading day." Finally, we derive the market-based average and volatility of return of different investors during the "trading day" as a function of volatilities and correlations of market trade values. That highlights the distribution of the "actual" return of market trade and can serve as a benchmark for "purchasing" investors.

This research received no support, specific grants, or financial assistance from funding agencies in the public, commercial, or nonprofit sectors. We welcome valuable offers of grants, support, and positions.

# 1. Introduction

The literature describes the two kinds of stock returns. For convenience, we note such returns as "anticipated" and "actual" returns. The most studied "anticipated" return $r(t,\tau)$ is determined by the ratio $r(t,\tau)=p(t)/p(t-\tau)$ of the stock price $p(t)$ traded "today" at time $t$ and the price $p(t-\tau)$ traded at $t-\tau$ in the past. The "anticipated" stock returns $r(t,\tau)$ describe the expected, anticipated gains or losses that investors could get if they bought stocks at time $t-\tau$ in the past and then sold them at time $t$ "today." Modeling and predictions of the "anticipated" stock return at horizon $T$ define the core issues of financial economics (Fisher and Lorie, 1964; Mandelbrot, Fisher, and Calvet, 1997; Campbell, 1985; Brown, 1989; Fama, 1990; Fama and French, 1992; Lettau and Ludvigson, 2003; Greenwood and Shleifer, 2013; van Binsbergen and Koijen, 2015; Martin and Wagner, 2019). The description of the probabilistic properties of the "anticipated" stock return during any specific averaging time interval $\varDelta$ "today," or, as we note, a "trading day," and predictions of the probability of return at horizon $T$ "next day," deliver the most desired results for investors and traders. The frequency-based analysis of the return time series assesses the probability distributions of the "anticipated" return during a "trading day" (Amaral et al., 2000; Andersen et al., 2001; Knight and Satchell, 2001; Tsay, 2005; Andersen and Benzoni, 2009).

However, the "anticipated" stock return $r(t,\tau)=p(t)/p(t-\tau)$ with the time shift $\tau$ describes the gains or losses that may or may not match the real, "actual" return of investors. As "actual," we consider the return that a particular investor gains from selling the stocks at the time $t$, which he has purchased in 10 minutes, a day, a week, or any time in the past. Obviously, not all stocks that investors sell at time $t$ "today" were purchased at time $t-\tau$ in the past. Some stocks were purchased earlier or later than $t-\tau$ and at a price that differs from $p(t-\tau)$. Thus, investors who sell stocks at time $t$ gain returns that are different from "anticipated" returns, $r(t,\tau)=p(t)/p(t-\tau)$. So, "anticipated" returns $r(t,\tau)$ describe an option investors may gain, and "actual" returns describe the benefits investors realize. During a "trading day," traders and investors sell stocks that were initially purchased at different times in the past. Investors sell stocks during the "trading day" and gain returns on stocks they purchased in 10 minutes, a week, or any time in the past. To assess the average returns, or statistical moments of "actual" returns, that investors gain, one should take into account returns with different time shifts. That differs "actual" return from the description of the statistical properties of "anticipated" return. Investigation of the "actual" returns of institutional, professional, or individual investors forms a separate problem. Different aspects of "actual" returns were studied by (Schlarbaum, Lewellen and Lease, 1978; Stanley, Lewellen and Schlarbaum, 1980; Baker and Wurgler, 2004; Ivković, Sialm and Weisbenner, 2004; Gabaix, et al 2005; Daniel and Hirshleifer, 2016; Koijen, Richmond and Yogo, 2020; Hardouvelis, Karalas and Vayanos, 2021) and others.

However, the statistical properties of "anticipated" and "actual" returns are mostly studied in the same way. To assess the probability *P(r)* of "anticipated" or "actual" returns during the time interval *Δ,* one studies the time series of returns and assesses the frequency $m_r/N$:

$$P(r) \sim \frac{m_r}{N} \qquad (1.1)$$

In (1.1), $m_r$ denotes the number of returns that equal *r,* and *N* is the total number of terms of the time series during *Δ*. That is the conventional assessment of the probability of any event, and analyzing its time series during *Δ* follows the basis of probability theory (Shephard, 1991; Shiryaev, 1999; Shreve, 2004). For convenience, we further refer to such assessments (1.1) as the frequency-based probabilities of stock returns. It is regular and completely correct to assess the probability of return if the time series of return $r(t_i,\tau)$ during the averaging interval *Δ* are considered independent, self-reliant variables. However, return at time $t_i$ is determined by stock price $p(t_i)$ at time $t_i$ and price $p(t_i-\tau)$ at time $t_i-\tau$. Moreover, market trade values $C(t_i)$, volumes $U(t_i)$, and price $p(t_i)$ at time $t_i$ obey the primitive trade pricing equation (1.2):

$$C(t_i) = p(t_i)U(t_i) \qquad (1.2)$$

The equation (1.2) states that the statistical properties of trade value $C(t_i)$ and volume $U(t_i)$ should determine the properties of price as a random variable during *Δ*. For convenience, in this paper, all prices are adjusted to the present time *t*. We consider market trade values, volumes, market prices, and returns as random variables during the selected time-averaging interval *Δ*. We believe that the consideration of market prices and stock returns as financial and economic matters should take into account the impact of the size of the trade values $C(t_i)$ and volumes $U(t_i)$ (1.2) on the average price, return, volatility, and statistics of returns. The well-known volume weighted average price (VWAP) (Berkowitz et al., 1988; Duffie and Dworczak, 2018), which differs from the frequency-based assessment of the mean price, demonstrates the impact of the size of trade volumes $U(t_i)$ on the average price. It is reasonable that the statistical properties of the market trade values $C(t_i)$ and volumes $U(t_i)$ determine the statistics of the market price, and in turn, the price statistics determine the statistics of stock return. The randomness of market trade determines the statistical moments of price and return, and that differs from the frequency-based probability assessments of the price and return time series. We denote as market-based, the dependence of the statistical moments of return on the statistical moments and correlations of market trade values $C(t_i)$ and volumes $U(t_i)$ during the interval $\Delta$. A description of the statistical moments of market prices and "anticipated" stock returns that depend on the statistical moments and correlations of market trade values $C(t_i)$ and volumes $U(t_i)$ is given in Olkhov (2021-2023). We use these references to describe the market-based statistical moments of "actual" return.

Our paper describes the market-based statistical moments of the "actual" returns of investors, which they gain during the averaging interval $\Delta$ that we denote a "trading day". We call all stocks that are sold by all investors during the averaging interval $\Delta$ a trading day portfolio. We derive the dependence of the statistical moments of the "actual" return of investors on the statistical moments and correlations of market trade values $C(t_i)$ and volumes $U(t_i)$. In turn, the statistical moments of market trade values and volumes are assessed by the regular frequency-based (1.1) probability (Shephard, 1991; Shiryaev, 1999; Shreve, 2004). We consider statistical moments of return, which all investors gain from the sale of the trading day portfolio, as benchmarks for investors who purchase stocks during the same "trading day".

In this paper we reduce our description by the first two market-based statistical moments of "actual" return and derive the dependence of average and volatility of return on statistical moments and correlations of market trade values and volumes for different cases.

In Section 2, we describe the dependence of the market-based averages and the volatilities of price and the "anticipated" stock return on statistical moments and correlations of the current and past trade values and volumes. In Section 3, we consider the dependence of market-based average and volatility of the "actual" return of a single deal as a result of numerous purchases in the past. Section 4 presents the average and volatility of the "actual" return, which an investor gains if he makes a lot of sales during the "trading day." In Section 5, we consider the market-based average and volatility of the return of different investors during the "trading day." Section 6, Conclusion. We assume that readers know the basics of probability theory, statistical moments, etc.

## 2. Market-based averages and volatilities of price and "anticipated" return

As "anticipated," we consider the stock return $r(t_i,\tau)=p(t_i)/p(t_i-\tau)$ determined as the ratio of market trade price $p(t_i)$ at time $t_i$ with respect to the price $p(t_i-\tau)$ at time $t_i-\tau$. We consider the trade of identical stocks and adjust all prices to the present. Let us consider the time series of the trade values $C(t_i)$ and volumes $U(t_i)$ at time $t_i$ and assume that the trades are made with a constant time shift $\varepsilon$:

$$t_{i+1} - t_i = \varepsilon \quad ; \quad \varepsilon - const$$

Market trade time series present irregular and highly variable data. To forecast the stock returns, one should choose the averaging interval $\Delta >> \varepsilon$ and estimate the average variables. We consider the present time $t=t_0$ as "today," and the time $t_k=t-k\Delta$, $k=1,2,..$ describes $k\Delta$ intervals in the past:

$$\Delta_k = \left[t_k - \frac{\Delta}{2}; t_k + \frac{\Delta}{2}\right] \quad ; \quad t_k = t - \Delta \cdot k \quad ; \quad k = 0, 1, 2, \ldots \qquad (2.1)$$

We assume that each interval $\Delta_k$, $k=0,1,..$, $\Delta_0 = \Delta$, contains the same number $N$ of terms $t_i$ of the time series:

$$t_i \in \Delta \quad ; \quad i = 1, 2, \ldots N \qquad (2.2)$$

We denote the averaging interval $\Delta$ (2.2) at present time $t$ as the "trading day." We consider the trade values $C(t_i)$, volumes $U(t_i)$, and prices $p(t_i)$ during each interval $\Delta_k$ (2.1) as random variables. To describe a random variable, one can equally use the probability measure, the characteristic function, or the set of statistical moments of the random variable (Shephard, 1991; Shiryaev, 1999; Shreve, 2004). Finite number of statistical moments describes approximations of the characteristic function and probability measure. We derive finite number of the statistical moments of price and return that describe approximations of their probability measures. We assess the statistical moments of the trade value $C(t;n)$ and volume $U(t;n)$ averaged during the "trading day" $\Delta$ (2.2) by finite number $N$ of trade time series using frequency-based probability (1.1) as:

$$C(t;n) = E[C^n(t_i)] \sim \frac{1}{N}\sum_{i=1}^{N} C^n(t_i) \qquad (2.3)$$

$$U(t;n) = E[U^n(t_i)] \sim \frac{1}{N}\sum_{i=1}^{N} U^n(t_i) \quad ; \quad n = 1,2,. \qquad (2.4)$$

We denote $E[\ldots]$ as the frequency-based mathematical expectation (2.3; 2.4) during $\Delta$ (2.2) and the symbol "~" to underline that the finite number $N$ of trades determines the assessments, the estimators of the statistical moments of trade value $C(t;n,)$ and volume $U(t;n)$ at time $t=t_0$ "today." The finite number $n$ of the statistical moments (2.3; 2.4) assesses only approximations of the probability and the characteristic functions of the random variables $C(t_i)$ and $U(t_i)$.

We denote $E_m[\ldots]$ as market-based mathematical expectation to differ it from the conventional frequency-based one (2.3; 2.4) and describe first two market-based statistical moments of price $a(t;n)=E_m[p^n(t_i)]$ and return $h(t,\tau;n)=E_m[r^n(t_i,\tau)]$ for $n=1,2$. We denote the "anticipated" return $r(t_i,\tau)$ (2.5) with time shift $\tau$ as the ratio of prices $p(t_i)$ at times $t_i$ to price $p(t_i-\tau)$ in the past:

$$r(t_i, \tau) = \frac{p(t_i)}{p(t_i - \tau)} \qquad (2.5)$$

Let us transform the trade price equation (1.2):

$$C(t_i) = p(t_i)U(t_i) = \frac{p(t_i)}{p(t_i-\tau)} p(t_i - \tau)U(t_i) = r(t_i, \tau)\, C_o(t_i, \tau)$$

Equations (2.6) link up the sale value $C(t_i)$, return $r(t_i,\tau)$, and the original value $C_o(t_i,\tau)$ of the same trade volume $U(t_i)$ at price $p(t_i-\tau)$ in the past:

$$C(t_i) = r(t_i, \tau)\, C_o(t_i, \tau) \quad ; \quad C_o(t_i, \tau) = p(t_i - \tau)U(t_i) \qquad (2.6)$$

Equation (2.6) is similar to the trade price equation (1.2), but the price $p(t_i-\tau)$ defines the original value $C_o(t_i,\tau)$ of the trade volume $U(t_i)$ in the past at time $t_i-\tau$. Similar to (2.3), we assess the frequency-based statistical moments $C_o(t,\tau;n)$ of the original value $C_o(t_i,\tau)$:

$$C_o(t, \tau; n) = E[C_o^n(t_i, \tau)] \sim \frac{1}{N}\sum_{i=1}^{N} C_o^n(t_i, \tau) \qquad (2.7)$$

The *m-th* degree of (1.2; 2.6) for $m=1,2,..$, give equations (2.8; 2.9):

$$C^m(t_i) = p^m(t_i)U^m(t_i) \qquad (2.8)$$

$$C_o^m(t_i, \tau) = p^m(t_i - \tau)U^m(t_i) \quad ; \quad C^m(t_i) = r^m(t_i, \tau)\, C_o^m(t_i, \tau) \qquad (2.9)$$

The equation (2.8) generates the set of weight functions $w(t_i;m)$ (2.10)

$$w(t_i; m) = \frac{U^m(t_i)}{\sum_{i=1}^{N} U^m(t_i)} \quad ; \quad \sum_{i=1}^{N} w(t_i; m) = 1 \qquad (2.10)$$

The weight functions $w(t_i;m)$ (2.10) define the average $p(t;n,m)$ (2.11) of the *n-th* degree of price $p^n(t_i)$:

$$p(t; n, m) = \sum_{i=1}^{N} p^n(t_i)w(t_i; m) = \frac{1}{\sum_{i=1}^{N} U^m(t_i)} \sum_{i=1}^{N} p^n(t_i)U^m(t_i) \qquad (2.11)$$

Relations (2.11) present generalization of the well-known volume weighted average price (VWAP) (Berkowitz et al., 1988; Duffie and Dworczak, 2018) $p(t;1,1)$ for $n=1,2,..$, and $m=1,2,..$ . We consider VWAP $p(t;1,1)$ as market-based average price $a(t;1)$ and take:

$$a(t; 1) = E_m[p(t_i)] = p(t; n, m) \qquad (2.12)$$

The choice (2.12) of market-based average price $a(t;1)$ determines the dependence of the first four market-based statistical moments $a(t;n)$, $n=2,3,4$ on statistical moments and correlations of trade value and volume (Olkhov, 2022). The dependence of the 2-d price statistical moment $a(t;2)$ (2.13) and price volatility $\sigma_p^2(t)$ (2.14) on the statistical moments and correlations of trade values and volumes take the form:

$$a(t; 2) = E_m[p^2(t_i)] = \frac{C(t;2) + 2a^2(t;1)\Omega_U^2(t) - 2a(t;1)corr\{C(t_i)U(t_i)\}}{U(t;2)} \qquad (2.13)$$

$$\sigma^2(t) = E_m\big[(p(t_i) - a(t; 1))^2\big] = \frac{\Omega_C^2(t) + a^2(t;1)\Omega_U^2(t) - 2a(t;1)corr\{C(t_i)U(t_i)\}}{U(t;2)} \qquad (2.14)$$

In (2.13; 2.14) $\Omega_C^2(t)$ (2.15) and $\Omega_U^2(t)$ (2.16) denote volatilities of trade value and volume:

$$\Omega_C^2(t) = E[(C(t_i) - C(t;1))^2] = C(t;2) - C^2(t;1) \qquad (2.15)$$

$$\Omega_U^2(t) = E[(U(t_i) - U(t;1))^2] = U(t;2) - U^2(t;1) \qquad (2.16)$$

The correlation $corr\{C(t_i)U(t_i)\}$ (2.17) between the trade value $C(t_i)$ and volume $U(t_i)$ is determined by the joint mathematical expectation $E[C(t_i)U(t_i)]$ of the trade value and volume:

$$CU(t;1,1) = E[C(t_i)U(t_i)] = C(t;1)U(t;1) + corr\{C(t_i)U(t_i)\} \qquad (2.17)$$

We refer to Olkhov (2021; 2022) for further details. The similar method allows the use of the equations (2.7; 2.9) to derive the dependence of market-based statistical moments of the "anticipated" returns on statistical moments and correlations of the current and past trade values and trade volumes (Olkhov, 2023). Similar to (2.11) the weight functions (2.18):

$$z(t,\tau;m) = \frac{C_o^m(t_i,\tau)}{\sum_{i=1}^N C_o^m(t_i,\tau)} \quad ; \quad \sum_{i=1}^N z(t_i,\tau;m) = 1 \qquad (2.18)$$

define the average $r(t,\tau;n,m)$ (2.19) of the *n-th* degree of return $r^n(t_i,\tau)$:

$$r(t,\tau;n,m) = \sum_{i=1}^N r^n(t_i,\tau)z(t,\tau;m) = \frac{1}{\sum_{i=1}^N C_o^m(t_i,\tau)} \sum_{i=1}^N r^n(t_i,\tau)C_o^m(t_i,\tau) \qquad (2.19)$$

The definition of the average return $r(t,\tau;1,1)$ (2.19) coincides with Markowitz's (1952) definition of the portfolio return as an average return weighted by the "relative amount $X_i$ invested in security *i*." To approve $r(t,\tau;1,1)$ (2.19), one can consider the shares sold during the "trading day" as a "portfolio" and use Markowitz's definition of portfolio return that was presented more than 35 years ahead of the definition of VWAP (Berkowitz et al., 1988). The replacement of "relative amount invested in security" by the relative volume of trade determined by the weight function $w(t_i;1)$ (2.10) determines the VWAP. We take Markowitz's definition of portfolio return in the form $r(t,\tau;1,1)$ (2.20) as market-based average return $h(t,\tau;1)$:

$$h(t,\tau;1) = E_m[r(t_i,\tau)] = r(t,\tau;1,1) \qquad (2.20)$$

In (Olkhov, 2023) we derive the dependence of the 2-d market-based statistical moment $h(t,\tau;2)$ (2.21) and return volatility $\sigma^2(t,\tau)$ (2.22) on statistical moments and correlations of the current and past trade values and volumes:

$$h(t,\tau;2) = E_m[r^2(t_i,\tau)] = \frac{C(t;2) + 2h^2(t,\tau;1)\Phi^2(t,\tau) - 2h(t,\tau;1)corr\{C(t_i)C_o(t_i,\tau)\}}{C_o(t,\tau;2)} \qquad (2.21)$$

$$\sigma^2(t,\tau) = E_m[(r(t_i,\tau) - h(t,\tau;1))^2] = \frac{\Omega_C^2(t) + h^2(t,\tau;1)\Phi^2(t,\tau) - 2h(t,\tau;1)corr\{C(t_i)C_o(t_i,\tau)\}}{C_o(t,\tau;2)} \qquad (2.22)$$

The *2-d* statistical moment $h(t,\tau;2)$ (2.21) and return volatility $\sigma^2(t,\tau)$ (2.22) depend on the volatility $\Omega_C^2(t)$ (2.15) of the current value and on volatility $\Phi^2(t,\tau)$ (2.23) of the past value:

$$\Phi^2(t,\tau) = E[(C_o(t_i,\tau) - C_o(t,\tau;1))^2] = C_o(t,\tau;2) - C_o^2(t,\tau;1) \qquad (2.23)$$

The joint mathematical expectations (2.24) determines the correlation $corr\{C(t_i)C_o(t_i,\tau)\}$ of current and past trade values:

$$E[C(t_i)C_o(t_i,\tau)] = \frac{1}{N}\sum_{i=1}^{N} C(t_i)C_o(t_i,\tau) = C(t;1)\,C_o(t,\tau;1) + corr\{C(t_i)C_o(t_i,\tau)\} \quad (2.24)$$

The relations (2.12-2.14; 2.20-2.22) describe the market-based average, the 2-d statistical moments, and volatilities of price and "anticipated" stock return, and we refer to Olkhov (2021-2023) for further details.

Now we use the above results to describe the market-based average and volatility of "actual" returns for three cases. In Section 3, for the stock return of a single sale, we assess the market-based average and volatility that are generated by numerous purchases of stocks in the past. In Section 4, we describe the market-based average and volatility of the return of a single investor, which he gains due to multiple sales during the "trading day." In Section 5, we describe the market-based volatility of returns that different investors gain during the "trading day."

## 3. Market-based "actual" return of a single sale

In this section, we consider market-based statistical moments of "actual" return that an investor gains within a single sale of the volume $U(t_i)$ of stocks. We propose that the investor, at time $t_i$ "today," sells $U(t_i)$ stocks at a price $p(t_i)$. We assume that the investor purchased this amount of stocks $U(t_i)$ by small stakes of shares $U(t_j(i))$ at different times $t_j(i)$, $j=1,2,..M(i)$ in the past at a price $p(t_j(i))$. We consider all prices $p(t_j(i))$ adjusted to the present at time $t$. The investor at time $t_j(i)$ purchases the original value $C_o(t_j(i))$ of the volume $U(t_j(i))$ of stocks at price $p(t_j(i))$:

$$C_o(t_j(i)) = p\big(t_j(i)\big)\,U\big(t_j(i)\big) \quad (3.1)$$

For each volume $U(t_j(i))$ of stocks purchased in the past at a price $p(t_j(i))$ we use equation (3.2) similar to (2.6):

$$C\big(t_i,t_j(i)\big) = p(t_i)U\big(t_j(i)\big) = \frac{p(t_i)}{p\big(t_j(i)\big)}p\big(t_j(i)\big)U\big(t_j(i)\big) = r\big(t_j(i)\big)\,C_o\big(t_j(i)\big)$$

$$r\big(t_i,t_j(i)\big) \equiv \frac{p(t_i)}{p\big(t_j(i)\big)}$$

$$C(t_i) = \sum_{j=1}^{M(i)} C\big(t_i,t_j(i)\big) \quad ; \quad C\big(t_i,t_j(i)\big) = r\big(t_i,t_j(i)\big)\,C_o\big(t_j(i)\big) \quad (3.2)$$

The equation (3.2) introduces the current $C(t_i,t_j(i))$ value of the small stakes of shares $U(t_j(i))$ that were originally purchased at price $p(t_j(i))$ in the past. At time $t_i$, the investor gains the "actual" return $r(t_i,t_j(i))$ by selling the volume $U(t_j(i))$ at price $p(t_i)$. We denote $C_o(t_j(i))$ (3.2) as the original value in the past at a price *p(t_j(i))* adjusted to the present. The total sale volume *U(t_i)* (3.3) was purchased in the past by *M(i)* small stakes of shares *U(t_j(i))* at prices *p(t_j(i))*. The original value *C_o(t_i)* (3.3) of the volume *U(t_i)* (3.4) takes the form:

$$C_o(t_i) = \sum_{j=1}^{M(i)} C_o(t_j(i)) = \sum_{j=1}^{M(i)} p\bigl(t_j(i)\bigr) U\bigl(t_j(i)\bigr) \tag{3.3}$$

$$U(t_i) = \sum_{j=1}^{M(i)} U\bigl(t_j(i)\bigr) \tag{3.4}$$

Obviously, the purchases of stocks in the past at different prices *p(t_j(i))* result in a different "actual" return *r(t_i,t_j(i))*. If the total number *M(i)* of the purchases is sufficiently large, then (3.2) allows derive the statistical moments of the "actual" returns *r(t_i,t_j(i))* of a single market sale at a time *t_i*. One should follow section 2 and, similar to (2.9), take the *m-th* power of (3.2):

$$C^m(t_i, t_j(i)) = r^m\bigl(t_i, t_j(i)\bigr) C_o^m\bigl(t_j(i)\bigr) \tag{3.5}$$

We introduce the *n-th* statistical moments *C(t_i;n)* (3.6) of the sale value *C(t_i,t_j(i))* at *t_i* and the n-th statistical moments *C_o(t_i;n)* (3.8) of the original value *C_p(t_j(i))* as:

$$C(t_i; n) = E[C^n(t_i, t_j(i))] \sim \frac{1}{M(i)} \sum_{j=1}^{M(i)} C^n(t_i, t_j(i)) \tag{3.6}$$

The average current value *C(t_i;1)* (3.6) for *n=1* multiplied by *M(i)* equals the current value *C(t_i)* (3.2) of the sale at time *t_i*:

$$C(t_i) = \sum_{j=1}^{M(i)} C(t_i, t_j(i)) = M(i) C(t_i; 1) \tag{3.7}$$

The statistical moments *C_o(t_i;n)* (3.8) of the original value *C_o(t_j(i))* take the form:

$$C_o(t_i; n) = E\left[C_o^n\bigl(t_j(i)\bigr)\right] \sim \frac{1}{M(i)} \sum_{j=1}^{M(i)} C_o^n(t_j(i)) \tag{3.8}$$

Similar to (2.18; 2.19), for the equations (3.5), obtain the weight functions *s(t_i,t_j(i));m)* (3.9):

$$s\bigl(t_i, t_j(i); m\bigr) = \frac{C_o^m\bigl(t_j(i)\bigr)}{\sum_{j=1}^{M(i)} C_o^m\bigl(t_j(i)\bigr)} \quad ; \quad \sum_{j=1}^{M(i)} s\bigl(t_i, t_j(i); m\bigr) = 1 \tag{3.9}$$

The weight functions *s(t_i,t_j(i));m)* (3.9) determine the average *r(t_i;n,m)* (3.10) of the *n-th* degree of return *r^n(t_i,t_j(i))* of a single deal:

$$r(t_i; n, m) = \sum_{j=1}^{M(i)} r^n\bigl(t_i, t_j(i)\bigr) s\bigl(t_i, t_j(i); m\bigr) \tag{3.10}$$

We follow (2.20) and define the dependence of the market-based average *g(t_i;1)* (3.11) "actual" return of a single sale of stocks at time *t_i* on the statistical moments of current and past trade values (3.6; 3.8; 3.10):

$$g(t_i; 1) = E_m\left[r\bigl(t_i, t_j(i)\bigr)\right] = r(t; 1,1) \tag{3.11}$$

We use (3.6; 3.8) and for the 2-d market-based statistical moment $g(t_i;2)$ (3.13) and volatility $\sigma_g^2(t_i)$ (3.14) of "actual" return of a single sale of stocks at time $t_i$ obtain relations (3.12-3.17) that are similar to (2.15; 2.21- 2.24):

$$g(t_i; 2) = E_m\left[r^2\left(t_i, t_j(i)\right)\right]$$

$$\sigma_g^2(t_i) = E_m\left[(r\left(t_i, t_j(i)\right) - g(t_i; 1))^2\right]$$

We define market-based volatility $\sigma_g^2(t_i)$ (3.13) as:

$$\sigma_g^2(t_i) = \sum_{j=1}^{M(i)} (r\left(t_i, t_j(i)\right) - g(t_i; 1))^2 s(t_i, t_j(i); 2) \qquad (3.12)$$

Equation (3.12) provide the consistency of the 1-st $g(t_i;1)$ (3.11) and 2-d $g(t_i;2)$ (3.13) market-based statistical moments of "actual" return of a single sale and guarantee non-negativity of volatility $\sigma_g^2(t_i)$ (3.14) of "actual" return:

$$g(t_i; 2) = \frac{C(t_i;2) + 2g^2(t_i;1)\Phi^2(t_i) - 2g(t_i;1)corr\{C(t_i,t_j(i))C_o(t_j(i))\}}{C_o(t_i;2)} \qquad (3.13)$$

$$\sigma_g^2(t_i) = \frac{\Omega_C^2(t_i) + g^2(t_i;1)\Phi^2(t_i) - 2g(t_i;1)corr\{C(t_i,t_j(i))C_o(t_j(i))\}}{C_o(t_i;2)} \qquad (3.14)$$

In (3.13; 3.14) we denote the volatilities of current $\Omega_C^2(t_i)$ (3.15) and past $\Phi^2(t_i)$ (3.16) trade values of a single sale at time $t_i$:

$$\Omega_C^2(t_i) = E[(C\left(t_i, t_j(i)\right) - C(t_i; 1))^2] = C(t_i; 2) - C^2(t_i; 1) \qquad (3.15)$$

$$\Phi^2(t_i) = E[(C_o\left(t_j(i)\right) - C_o(t_i; 1))^2] = C_o(t_i; 2) - C_o^2(t_i; 1) \qquad (3.16)$$

The relations (3.17; 3.18) determine the correlation $corr\{C(t_i,t_j(i))C_o(t_j(i))\}$ between the current $C(t_i,t_j(i))$ trade value at $t_i$ of stocks that were originally purchased in the past at $t_j(i)$ and their original value $C_o(t_j(i))$:

$$CC_o(t_i) = E\left[C\left(t_i, t_j(i)\right)C_o\left(t_j(i)\right)\right] = \frac{1}{M(i)}\sum_{j=1}^{M(i)} C\left(t_i, t_j(i)\right)C_o\left(t_j(i)\right) \qquad (3.17)$$

$$CC_o(t_i) = C(t_i; 1)C_o(t_i; 1) + corr\left\{C\left(t_i, t_j(i)\right)C_o\left(t_j(i)\right)\right\} \qquad (3.18)$$

## 4. Market-based "actual" return of a single investor

In this section, we consider the market-based average and volatility of the "actual" return of a single investor during the "trading day" (2.2).

Actually, different deals at times $t_i$, $i=1,2,..N$ during the "trading day" (2.2) result in different market-based average returns $g(t_i;n)$ (3.11). One can consider different returns $g(t_i;n)$ (3.11) that a single investor gains during the "trading day" as a random variable. To describe the market-based random properties of average returns $g(t_i;n)$ (3.11) we consider equation (4.1) that describe the dependence of on market-based average return $g(t_i;1)$ of a single deal during the "trading day" on average current $C(t_i;1)$ (3.6) and past $C_o(t_i;1)$ (3.8) trade values:

$$C(t_i;1) = g(t_i;1)C_o(t_i;1) \tag{4.1}$$

In the equation (4.1) we consider the current $C(t_i;1)$ (3.6), past $C_o(t_i;1)$ (3.8) trade values, and average return $g(t_i;1)$ at time $t_i$ as random variables during the "trading day" (2.2). One can consider (4.1) similar to the trade return equations (3.2; 3.5) and reproduce the same calculations but with respect to the market-based average return $g(t_i;1)$ (3.11) of a single deal. Let us take the n-th degree of (4.1):

$$C^n(t_i;1) = g^n(t_i;1)C_o^n(t_i;1) \tag{4.2}$$

Similar to (3.6-3.10) we define the frequency-based n-th statistical moments (4.3; 4.4) of current and past trade values:

$$C(t;1|n) = E[C^n(t_i;1)] \sim \frac{1}{N}\sum_{i=1}^{N} C^n(t_i;1) \tag{4.3}$$

$$C_o(t;1|n) = E[C_o^n(t_i;1)] \sim \frac{1}{N}\sum_{i=1}^{N} C_o^n(t_i;1) \tag{4.4}$$

Similar to (3.9), we define the weight functions $\gamma(t_i;1|m)$ (4.5):

$$\gamma(t_i;1|m) = \frac{C_o^m(t_i;1)}{\sum_{j=1}^{M(i)} C_o^m((t_i;1))} \quad ; \quad \sum_{i=1}^{N} \gamma(t_i;1|m) = 1 \tag{4.5}$$

The weight functions $\gamma(t_i;1|m)$ (4.5) and the equation (4.2) define the average $\varrho(t;1|n,m)=E[g^n(t_i;1)]$ (4.6) of the n-th degree $g^n(t_i;1)$:

$$\varrho(t;1|n,m) = \sum_{j=1}^{M(i)} g^n(t_i;1)\,\gamma(t_i;1|m) \tag{4.6}$$

Similar to (3.11) we choose the market-based average of return $G(t|1)$ (4.7) that is determined by numerous deals of a single investor during the "trading day" to be equal the average $\varrho(t;1|1,1)$ (4.6):

$$G(t|1) = E_m[g(t_i;1)] = \varrho(t;1|1,1) \tag{4.7}$$

Similar to (3.12-3.17) the choice of $G(t|1)$ (4.7) determines the market-based 2-d statistical moment $G(t|2)$ (4.8) and the volatility $\sigma_G^2(t)$ (4.9) of the return of a sing investor:

$$G(t|2) = E_m[g^2(t_i;1)]$$

$$G(t;1|2) = \frac{C(t;1|2) + 2G^2(t|1)\Phi^2(t) - 2G(t|1)\,corr\{C(t_i;1)C_o(t_i;1)\}}{C_o(t;1|2)} \tag{4.8}$$

$$\sigma_G^2(t_i) = E_m[(g(t_i;1) - G(t;1|1))^2]$$

$$\sigma_G^2(t) = \frac{\Omega_C^2(t) + G^2(t|1)\Phi^2(t) - 2G(t|1)\,corr\{C(t_i;1)C_o(t_i;1)\}}{C_o(t;1|2)} \tag{4.9}$$

The market-based volatility $\sigma_G^2(t)$ (4.9) of return of a single investor depends on volatility $\Omega_C^2(t)$ (4.10) of his average current sales $C(t_i;1)$ (3.6) and on volatility $\Phi^2(t)$ (4.11) of the average original values $C_o(t_i;1)$ (3.8) at times $t_i$ during the "trading day":

$$\Omega_C^2(t) = E[(C(t_i; 1) - C(t; 1|1))^2] = C(t; 1|2) - C^2(t; 1|1) \tag{4.10}$$

$$\Phi^2(t) = E[(C_o(t_i; 1) - C_o(t; 1|1))^2] = C_o(t; 1|2) - C_o^2(t; 1|1) \tag{4.11}$$

As well, the market-based volatility $\sigma_G^2(t)$ (4.9) depends on correlation $corr\{C(t_i;1)C_o(t_i;1)\}$ (4.13) between the average current sales $C(t_i;1)$ (3.6) and original values $C_o(t_i;1)$ (3.8) at times $t_i$ during the "trading day". From (4.3; 4.4) obtain:

$$E[C(t_i; 1)C_o(t_i; 1)] = \frac{1}{N}\sum_{i=1}^{N} C(t_i; 1)C_o(t_i; 1) \tag{4.12}$$

$$E[C(t_i; 1)C_o(t_i; 1)] = C(t; 1|1)C_o(t; 1|1) + corr\{C(t_i; 1)C_o(t_i; 1)\} \tag{4.13}$$

## 5. Market based volatility of return of different investors

In this section, we assess the market-based average and volatility of "actual" return that many different investors gain during the "trading day." Indeed, the average return that each of numerous investors gain during the "trading day" varies a lot. One can consider the average returns of many different investors as a random variable. The market-based volatility of average returns describes the uncertainty of trade outcome of investors during the "trading day."

Let us assume that during the "trading day" the investor $q$, $q=1,..Q$ gain average return $G(t;1|q)$ (4.7) and consider the trade return equation (5.1) on average current value $C(t;1|1,q)$ (4.3), original value $C_0(t;1|1,q)$, and average return $G(t;1|q)$ (4.7) of investor $q$:

$$C(t; 1|1, q) = G(t; 1|q) \, C_o(t; 1|1, q) \quad ; \quad q = 1, \dots Q \tag{5.1}$$

In (5.1) we add variable q in the current value $C(t;1|1,q)$ (4.3), original value $C_0(t;1|1,q)$, and average return $G(t;1|q)$ (4.7) to highlight their dependence on trade outcomes of investor q during the "trading day." We consider the equation (5.1) similar to (4.1) and take the n-th degree of (5.1):

$$C^n(t; 1|1, q) = G^n(t; 1|q) \, C_o^n(t; 1|1, q) \tag{5.2}$$

Similar to (4.3; 4.4) we define the frequency-based m-th statistical moments of current $C(t|m)$ (5.3) and past $C_o(t|m)$ (5.4) trade values that are determined by of different investors during the "trading day":

$$C(t|m) = \frac{1}{Q}\sum_{q=1}^{Q} C^m(t; 1|1, q) \tag{5.3}$$

$$C_o(t|m) = \sum_{q=1}^{Q} C_o^m(t; 1|1, q) \tag{5.4}$$

Similar to (4.5), equation (5.2) defines the weight functions $\varphi(t|m,q)$ (5.5):

$$\varphi(t|m, q) = \frac{C_o^m(t;1|1,q)}{\sum_{q=1}^{Q} C_o^m(t;1|1,q)} \quad ; \quad \sum_{q=1}^{Q} \varphi(t|m, q) = 1 \tag{5.5}$$

The weight functions $\varphi(t|m,q)$ (5.5) define the frequency-based averages $D(t|n,m)$ (5.6) of the n-th degree of return $G^n(t;1|q)$ (4.7) over the set of numerous investors $q=1,2,..Q$:

$$D(t|n,m) = \sum_{q=1}^{Q} G^n(t;1|q)\,\varphi(t|m,q) \tag{5.6}$$

We define the market-based average return R(t|1) (5.7) as:

$$R(t|1) = E_m[G(t_i;1|q)] = D(t|1,1) \tag{5.7}$$

Similar to (4.9) we define market-based volatility $\sigma_R^2(t)$ (5.8) of return determined by numerous different investors during the "trading day":

$$\sigma_R^2(t_i) = E_m[(G(t_i;1|q) - R(t|1))^2]$$

$$\sigma_R^2(t_i) = \sum_{q=1}^{Q} (G(t_i;1|q) - R(t|1))^2\,\varphi(t|m,q)$$

$$\sigma_R^2(t) = \frac{\Omega_R^2(t) + R^2(t|1)\Phi_R^{\,2}(t) - 2R(t|1)\,corr\{C((t;1|1,q))C_o((t;1|1,q))\}}{C_o(t|2)} \tag{5.8}$$

The market-based volatility $\sigma_R^2(t)$ (5.8) depends upon the volatilities of current $\Omega_R^2(t)$ (5.9) and past $\Phi_R^{\,2}(t)$ (5.10) trade values during the "trading day":

$$\Omega_R^2(t) = E[(C(t;1|1,q) - C(t|1))^2] = C(t|2) - C^2(t|1) \tag{5.9}$$

$$\Phi_R^{\,2}(t) = E[(C_o(t;1|1,q) - C_o(t|1))^2] = C_o(t|2) - C_o^2(t|1) \tag{5.10}$$

The market-based volatility $\sigma_R^2(t)$ (5.8) also depends on correlation $corr\{C(t;1|1,q)C_o(t;1|1,q)\}$ (5.12) between the current $C(t;1|1,q)$ (5.3) and original values $C_0(t;1|1,q)$ (5.4) of investors $q=1,2,..Q$ during the "trading day":

$$E[C((t;1|1,q))C_o((t;1|1,q))] = \frac{1}{Q}\sum_{q=1}^{Q} C((t;1|1,q))C_o((t;1|1,q)) \tag{5.11}$$

$$E[C((t;1|1,q))C_o((t;1|1,q))] = C(t|1)C_o(t|1) + corr\{C((t;1|1,q))C_o((t;1|1,q))\} \tag{5.12}$$

## 6. Conclusion

This paper describes three successive approximations of the market-based averages and volatilities of the "actual" return that the investors gain during the "trading day." We describe the approximations of return generated by a single trade sale, by a single investor, and by all investors during the "trading day." We derive the dependence of the market-based averages and volatilities on statistical moments, volatilities, and correlations of the current and past trade values. Let us highlight some problems that seem to be important for the description of financial markets.

The market-based average and volatility of return that all investors "actually" gain as a result of their sales during the "trading day" can serve as benchmarks and impact the decisions of "purchasing" investors in financial markets. "Purchasing" investors can assess their forecast of the expected returns at horizon *T* in comparison with the "actual" returns that "selling" investors already gain. Analysis of relations between the statistical moments of return that "selling" investors already gain and predictions of the statistical moments of return of "purchasing" investors can help develop further asset pricing models and portfolio theory. The volatility of return (5.8) describes the distribution of the "actual" returns over numerous investors in the stock market.

Probably, it is difficult to collect and study the market data that permit the assessments of market-based averages and volatilities of the "actual" return of a single investor and of all investors during the "trading day." The market data that define the market-based averages and volatilities of "anticipated" returns (Section 2) are much more available. It is important to study relations between statistical moments of "anticipated" and "actual" return and highlight possible dependence between these factors.

The fluctuations of the "anticipated" returns due to the variations of the time shift $\tau$ can impact the duration of stock holding by the investors. That, in turn, can change the scales and fluctuations of "anticipated" returns determined by the time shift $\tau$. Investigation of the hidden mutual dependence between the market-based statistics of the "actual" return of investors and the statistics of the "anticipated" return can help increase the efficiency of portfolio performance and asset pricing models.